\documentclass[11pt]{article}

\usepackage[a4paper,margin=1in]{geometry}
\usepackage[T1]{fontenc}
\usepackage{amsmath,amssymb}
\usepackage{bm}
\usepackage{graphicx}
\usepackage{xcolor}
\usepackage[normalem]{ulem}

\begin{document}

\title{Heat Transport of the $\beta$-Fermi--Pasta--Ulam--Tsingou chain in the long-wave limit}

\author{Henrique Santos Lima$^{1}$\\
Matheus M. R. Poltronieri Martins$^{1}$\\
Lucas Reis e Silva$^{2}$\\[0.5em]
\small $^{1}$ Centro Brasileiro de Pesquisas F\'isicas, Rua Xavier Sigaud 150,
Rio de Janeiro 22290-180, Brazil\\
$^{2}$ Universidade Federal de Santa Catarina,\\
Campus Universit\'ario Reitor Jo\~ao David Ferreira Lima,\\
Trindade, Florian\'opolis, Santa Catarina 88040-900, Brazil}

\date{\today}

\maketitle

\begin{abstract}
Resolving the anomalous conductivity exponent of the symmetric
$\beta$-Fermi--Pasta--Ulam--Tsingou chain by molecular dynamics can require
very large systems because of long finite-size crossovers and thermal-contact
resistance. Motivated by this computational challenge, we derive a
long-wavelength continuum description and investigate whether the
kinetic-theory scaling $\kappa\propto L^{2/5}$ becomes accessible with a moderate
number of numerical degrees of freedom. The nonlinear elastic field retains
the cubic stress of the microscopic interaction and exchanges heat with
Langevin reservoirs. A flux-conservative spatial discretization constructs
the force and energy current from the same stress, providing a consistent
interior transport estimator. For $L>8$, corresponding to approximately
$10^3$ mesh nodes and above at the reference resolution, the fitted exponent
is $0.399\pm0.004$ . Mesh refinement supports
the stability of this exponent between the two finer resolutions, although
the conductivity amplitude remains resolution-dependent. The main result
is thus an accessible finite-size regime near $2/5$, rather than convergence
of the absolute conductivity as the mesh spacing vanishes. This formulation
provides a practical route to studying anomalous transport in anharmonic
systems and a conservative continuum framework for investigating energy
flow in nonlinear elastic media.
\end{abstract}

\section{Introduction}

Fourier's law is one of the central constitutive relations of macroscopic
transport: a temperature gradient produces a heat current with a material
coefficient that remains finite in the thermodynamic limit.  In one spatial
dimension, however, this apparently simple statement becomes a fundamental
problem of nonequilibrium statistical mechanics.  Reduced dimensionality
enhances long-wavelength fluctuations, and the simultaneous conservation of
energy and momentum can prevent current correlations from decaying rapidly
enough to yield a finite bulk conductivity.  The resulting transport is
anomalous: the conductivity depends on system size, energy spreads
superdiffusively, and boundary and bulk effects remain coupled over unusually
long scales \cite{prosen2000,narayan2002,lepri2003,dhar2008}.

At sufficiently large scales, the conserved fields and their nonlinear
couplings, rather than microscopic details of the force law, determine the
transport behavior. Nonlinear fluctuating hydrodynamics organizes this regime
through interacting sound and heat modes \cite{vanbeijeren2012,spohn2014}.

The Fermi--Pasta--Ulam--Tsingou (FPUT) chain is a canonical setting for testing
this connection between Hamiltonian microscopic dynamics and macroscopic
transport \cite{fput1955,mai2007}. Its reflection-symmetric $\beta$ variant
isolates quartic phonon interactions. In the weakly anharmonic regime,
phonon-kinetic theory predicts a $t^{-3/5}$ current-correlation tail and
$\kappa\sim L^{2/5}$ \cite{pereverzev2003,lukkarinen2008}. Extracting
this regime numerically is nevertheless delicate because contact resistance
and long crossovers can mask the intrinsic bulk scaling
\cite{takatsu2024,lepri2026}. The kinetic-theory scaling serves
as a benchmark for the finite-size behavior examined here.

Recent wave-kinetic simulations connect this crossover to the coexistence of
nearly ballistic low-wavenumber modes and relaxing kinetic modes
\cite{dematteis2020,devita2022}; boundary resistance can further affect the
apparent conductivity \cite{sun2023}. These observations motivate a numerical
formulation in which the discrete force and the measured energy current are
constructed from the same stress flux.

In the present construction, the same bond stress determines both
the force on each node and the energy transferred across each bond.
For the conservative semidiscrete dynamics, changes in the local
total energy are balanced by the energy currents through neighboring bonds.
This avoids using a current definition that is
inconsistent with the equation of motion and makes the separation between
interior transport and reservoir contacts more transparent, providing a
consistent basis for interpreting the current plateau and the conductivity
scaling observed here.

In continuum mechanics, the $\beta$-FPUT
interaction is the lowest-order reflection-symmetric correction to Hookean
elasticity: a quartic strain-energy density produces a cubic, reversible
stress. This conservative constitutive form provides a minimal longitudinal
hyperelastic model for nonlinear wave propagation and
energy transport \cite{fuogden2001,hiki1966,barth2024,zhang2024}.

We provide a continuum-to-lattice implementation in which the force
discretization and the
heat-current observable are compatible by construction. The practical question
is whether this construction makes anomalous scaling visible at moderate system
sizes, outside the short-length contact-dominated range. We therefore keep
the spatial resolution fixed, resolve the stress on bonds, and extract the
conductivity from an interior temperature drop that excludes the thermostat
contact region. The results identify the onset of this regime within the
present formulation.

The nodes represent a numerical discretization of the continuum
field, not microscopic particles of the original chain; we discuss the
implications for finite-size scaling and mesh dependence with the results.

\begin{figure}[!t]
    \centering
    \includegraphics[width=0.95\textwidth]{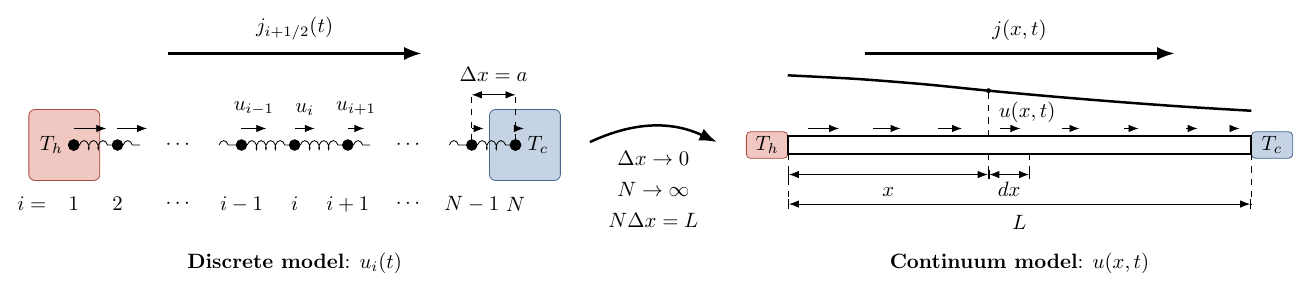}
    \caption{Long-wave passage from the discrete $\beta$-FPUT chain to the
    continuum displacement field $u(x,t)$.  When the lattice spacing
    $\Delta x$ is small compared with the length scale over which the
    displacement varies, the bond extension per unit spacing approaches
    the continuum strain,
    $(u_{i+1}-u_i)/\Delta x\to\partial_x u$ and the bond current becomes the
    smooth flux $j(x,t)$.  The hot and cold reservoirs remain boundary
    sources and sinks of energy in this limit.}
    \label{fig:continuum-limit}
\end{figure}

Figure~\ref{fig:continuum-limit} illustrates the passage from the
discrete chain to the continuum description used here. At length scales much
larger than the lattice spacing, the extension of individual bonds is
represented by a smooth strain field, while the quartic interaction becomes a
nonlinear elastic stress 
characterized by an effective $\beta$, namely
$\beta_{\text{eff}}$. The thermal
reservoirs remain localized at the two ends, where they act as boundary sources
and sinks of energy. The continuum model therefore preserves the long-wave
conservative dynamics of the $\beta$-FPUT chain while providing the
description employed in the simulations below.

Here we formulate the $\beta$-FPUT dynamics as a nonlinear elastic field
coupled at its ends to Langevin reservoirs. The continuum and lattice
descriptions are constructed around the same conservative stress flux, and the
heat current is measured on the corresponding interfaces. This provides a
controlled basis for comparing temperature profiles and currents across system
sizes and for testing anomalous finite-size scaling.

\section{Model and methods}

\subsection{Discrete chain and long-wave limit}

Let $x_n(t)$ be the displacement of the $n$th mass from its
equilibrium position, let $p_n=m\dot{x}_n$, and let $a$ be the equilibrium
lattice spacing. Define the extension of bond $(n,n+1)$ by
$r_n=x_{n+1}-x_n$. Each interior mass is coupled to its two nearest neighbors,
$n-1$ and $n+1$. Excluding the reservoir forces, the Hamiltonian is
\begin{equation}
    H_{\mathrm d}=
    \sum_{n=0}^{N-1}\frac{p_n^2}{2m}
    +\sum_{n=0}^{N-2}\left[
    \frac{K}{2}r_n^2+\frac{\beta}{4}r_n^4\right].
    \label{eq:discrete-hamiltonian}
\end{equation}
The term with index $n$ in the second sum is the energy of the
bond joining masses $n$ and $n+1$. Consequently, an interior displacement
$x_n$ occurs in two bond energies: the $(n-1,n)$ bond through
$r_{n-1}=x_n-x_{n-1}$ and the $(n,n+1)$ bond through
$r_n=x_{n+1}-x_n$. The two
nearest-neighbor interactions are therefore already contained in the
Hamiltonian without counting either bond twice. Here $K$ and $\beta$ are the
harmonic and quartic bond coefficients of the discrete $\beta$-FPUT chain. The two nearest-neighbor contributions are explicit
in the interior equation of motion,
\begin{equation}
    \begin{aligned}
    m\ddot{x}_n={}&K\left(r_n-r_{n-1}\right)
    +\beta\left(r_n^3-r_{n-1}^3\right).
    \end{aligned}
    \label{eq:discrete-eom}
\end{equation}
The continuum description is the leading long-wave
approximation to this nearest-neighbor interaction. To identify its bulk form,
write the displacement in Fourier modes. The discrete difference then becomes
\begin{equation}
    r_n=
    \frac{1}{\sqrt{N}}\sum_k
    \left(e^{\mathrm{i}ka}-1\right)x_k e^{\mathrm{i}kna}.
    \label{eq:fourier-difference}
\end{equation}
In the uniform interior of a long chain, the quartic part of
Eq.~\eqref{eq:discrete-hamiltonian} contains the four-mode factor
$\prod_{r=1}^{4}(e^{\mathrm{i}k_r a}-1)$, with the mode sum restricted by
wave-number conservation. For $|k_r a|\ll1$, each difference factor is
$e^{\mathrm{i}k_r a}-1\simeq\mathrm{i}k_r a$. The inverse Fourier transform
therefore gives a local quartic contribution proportional to
$(\partial_x u)^4$; the harmonic term similarly gives
$(\partial_x u)^2$. Identifying $u(na,t)=x_n(t)$, introducing a transverse
cross-sectional area $A$, and replacing the bond sum by an integral yields, to
leading long-wave order,
\begin{equation}
    \rho=\frac{m}{Aa},\qquad
    Y=\frac{Ka}{A},\qquad
    \beta_{\text{eff}}\equiv\frac{\beta a^3}{A}.
    \label{eq:continuum-parameters}
\end{equation}
Let us emphasize that $x_n(t)$ denotes the displacement from the
equilibrium position, not the absolute position of the $n$th mass. This is a
choice of reference coordinates, rather than an additional physical
assumption. If the absolute position is denoted by $X_n(t)$, then
$X_n(t)=na+x_n(t)$, and the extension of the $(n,n+1)$ bond relative to its
equilibrium length is
$X_{n+1}-X_n-a=x_{n+1}-x_n=r_n$. The displacement representation therefore
makes the undeformed chain the zero-strain reference state. The long-wave
approximation enters separately when $r_n/a$ is represented by
$\partial_x u$, with the continuum parameters defined relative to this
reference state.

\subsection{Continuum model}

We consider a one-dimensional nonlinear elastic field on $x\in[0,L]$ with longitudinal velocity $v=\partial_t u$ and strain $\varepsilon=\partial_x u$.  The constitutive stress is written as
\begin{equation}
    \sigma(\varepsilon)=Y\varepsilon+\beta_{\text{eff}}\varepsilon^3,
\end{equation}
where $Y$ is the Young's modulus, $Y\varepsilon$ is the linear elastic
contribution, and $\beta_{\text{eff}}\varepsilon^3$ is the nonlinear elastic, or
anharmonic, contribution.  Since the stress is the derivative of the elastic
energy density with respect to strain, the strain-energy density is
\begin{equation}
    W(\varepsilon)=\int_0^\varepsilon \sigma(\varepsilon')\,d\varepsilon'
    =\frac{Y}{2}\varepsilon^2+\frac{\beta_{\text{eff}}}{4}\varepsilon^4.
\end{equation}
The constitutive law is hyperelastic and reversible: the stress follows from
the single-valued potential $W(\varepsilon)$
\cite{fuogden2001,mihai2017}. Thus, the Hamiltonian $\beta$-FPUT model contains
no bulk dissipative variables; the Langevin damping and noise introduced below
describe energy exchange only at the boundaries.

For the hardening case considered here, $\beta_{\text{eff}}>0$, the tangent modulus
\begin{equation}
    \frac{d\sigma}{d\varepsilon}=Y+3\beta_{\text{eff}}\varepsilon^2,
\end{equation}
increases with the strain amplitude.  Thus, the bonds become progressively
stiffer while remaining reversible.  In continuum experiments, analogous
stress-dependent tangent moduli are detected through acoustoelastic shifts of
wave velocity \cite{hiki1966,barth2024,zhang2024}.

In the terminology of higher-order elasticity, the quartic energy term
$\beta_{\text{eff}}\varepsilon^4/4$ is the one-dimensional analogue of a
fourth-order elastic contribution and generates a cubic stress. The reflection
symmetry $W(-\varepsilon)=W(\varepsilon)$ removes the cubic energy term, making
the quartic term the leading nonlinear correction in the $\beta$-FPUT model
\cite{clayton2025,telyatnik2024}.

The bulk Hamiltonian is therefore
\begin{equation}
    H=A\int_0^L dx\left[\frac{\rho}{2}(\partial_t u)^2
    +\frac{Y}{2}(\partial_x u)^2
    +\frac{\beta_{\text{eff}}}{4}(\partial_x u)^4\right],
\end{equation}
where $A$ is the transverse cross-sectional area of the bar and $\rho$ is the volume mass density.
Hamilton's principle applied to the corresponding Lagrangian density gives
$\rho\partial_t^2u=\partial_x(\partial W/\partial\varepsilon)$.  Using
$\varepsilon=\partial_x u$ and $\partial W/\partial\varepsilon=\sigma$
therefore yields the bulk equation of motion
\begin{equation}
\begin{split}
    \rho\partial_t^2 u&=\partial_x\sigma\\
    &=\left[Y+3\beta_{\text{eff}}(\partial_x u)^2\right]\partial_x^2u.
\end{split}
    \label{eq:bulk}
\end{equation}
In the harmonic limit $\beta_{\text{eff}}=0$, the normal modes of an ideal chain are
independent and propagate without intrinsic phonon--phonon scattering,
producing ballistic transport.  The quartic energy for $\beta_{\text{eff}}\ne0$ couples
these modes and enables energy exchange through four-phonon interactions.
This anharmonic scattering promotes thermalization but does not destroy the
Hamiltonian conservation laws.  Momentum and energy conservation therefore
leave long-wavelength hydrodynamic modes long lived, allowing the transport to
remain anomalous rather than becoming conventionally diffusive
\cite{katsnelson2004,lukkarinen2008}.

In an idealized endpoint coupling, the boundary traction is set by Langevin reservoirs at
temperatures $T_h$ and $T_c$. Writing $\gamma_0$ as a boundary friction per
unit area, the force-balance conditions are
\begin{align}
    \sigma(\partial_x u(0,t))
    &=\gamma_0\partial_t u(0,t)
      -\sqrt{\frac{2\gamma_0 k_{\mathrm B}T_h}{A}}\,\xi_h(t),\\
    \sigma(\partial_x u(L,t))
    &=-\gamma_0\partial_t u(L,t)
      +\sqrt{\frac{2\gamma_0 k_{\mathrm B}T_c}{A}}\,\xi_c(t),
\end{align}
where $\xi_h$ and $\xi_c$ are independent Gaussian white-noise processes with
\begin{equation}
\begin{gathered}
    \langle \xi_a(t)\rangle=0,\\
    \langle \xi_a(t)\xi_b(t')\rangle=\delta_{ab}\delta(t-t').
\end{gathered}
\end{equation}

\subsection{Reduced units and numerical parameters}

The simulations implement the reservoirs as finite layers of
thermostatted nodes with free mechanical end conditions, rather than as
pointwise boundary tractions. The layer width and coupling strength therefore
form part of the numerical reservoir specification.

To connect the continuum coefficients to the quantities displayed in the
figures, we introduce a reference length $\ell_0$, the elastic wave speed
$c_0=\sqrt{Y/\rho}$, the time scale $t_0=\ell_0/c_0$, and the energy scale
$E_0=AY\ell_0$.  The reduced variables are
\begin{equation}
\begin{gathered}
    \widetilde{x}=\frac{x}{\ell_0},\qquad
    \widetilde{u}=\frac{u}{\ell_0},\qquad
    \widetilde{t}=\frac{t}{t_0},\\
    \widetilde{\sigma}=\frac{\sigma}{Y},\qquad
    \widetilde{\beta}_{\text{eff}}=\frac{\beta_{\text{eff}}}{Y},\qquad
    \widetilde{T}=\frac{k_{\mathrm B}T}{E_0},\\
    \widetilde{\gamma}_0=\frac{\gamma_0}{\sqrt{\rho Y}},
    \qquad \widetilde{H}=\frac{H}{E_0}.
\end{gathered}
\label{eq:reduced-units}
\end{equation}
In these variables the bulk equation has unit linear elastic coefficient and
depends on the anharmonicity only through $\widetilde{\beta}_{\text{eff}}$.  The numerical
results are reported in reduced units with
$\ell_0=A=\rho=k_{\mathrm B}=Y=1$; tildes are consequently omitted in the
simulation equations. Figure headings use $\widetilde{\beta}$ as a shorthand
for $\widetilde{\beta}_{\text{eff}}$; because $Y=1$, it has the same numerical
value as the $\beta_{\text{eff}}$ notation in Fig.~\ref{fig:kappa}. The
temperatures $T_h$ and $T_c$ are reduced in the same way, and
$T_0=(T_h+T_c)/2$ denotes their mean.

\begin{table}[t]
\caption{Numerical parameters for the reference resolution
$\Delta x=1/127$. The mesh-refinement comparison is described in
Sec.~\ref{sec:mesh-refinement}.}
\label{tab:numerical-parameters}
\begin{tabular}{lc}
\hline
Parameter & Value\\
Reduced constants $Y=\rho=A=k_{\mathrm B}$ & $1$\\
Anharmonicity $\widetilde{\beta}_{\text{eff}}=\beta_{\text{eff}}/Y$ & $0.5$\\
Reservoir temperatures $(T_h,T_c)$ & $(142.5,57.5)$\\
Spatial spacing $\Delta x$ & $1/127$\\
Time step $\Delta t$ & $0.001\,\Delta x$\\
Thermostat strength $\gamma\Delta t$ & $0.005$\\
Thermostatted nodes at each end $b$ & $4$\\
Ensemble size $M$ & $128$\\
Thermalization steps $n_{\mathrm{th}}$ & $2.0\times10^6(N-1)/127$\\
Production steps $n_{\mathrm{prod}}$ & $2.0\times10^5(N-1)/127$\\
\hline
\end{tabular}
\end{table}

\begin{figure}[!t]
    \centering
    \includegraphics[width=0.95\textwidth]{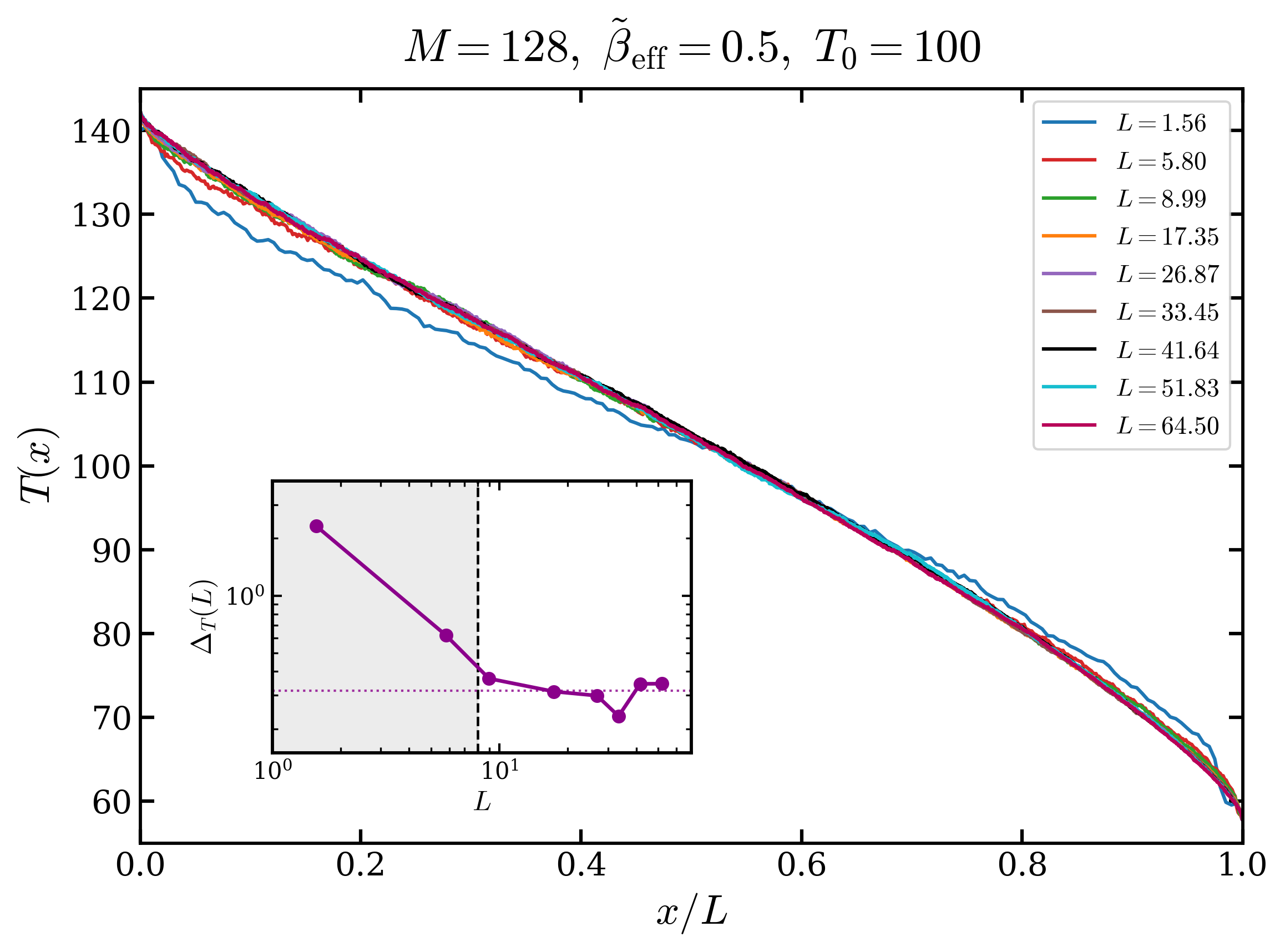}
    \caption{Stationary kinetic-temperature profiles for several lengths,
    plotted against the reduced coordinate $x/L$.  The temperature decreases
    monotonically from the hot to the cold reservoir, while the near-collapse
    of the curves for $L>8$ shows that the macroscopic profile is approximately
    scale-invariant. The inset quantifies this collapse through the
    root-mean-square distance $\Delta_T(L)$ from the $L_{\max}=64.50$ reference
    profile over $0.1\leq x/L\leq0.9$ and provides a profile-based criterion
    for the scaling range used below.}
    \label{fig:temperature-profile}
\end{figure}

In the numerical simulations, the interval contains $N$ nodes and
$N_b=N-1$ bonds, with
\begin{equation}
    \Delta x=\frac{L}{N_b}=\frac{L}{N-1}.
\end{equation}
For the reference data set, the spacing $\Delta x=1/127$ is fixed
for all lengths, so $N$ increases with
$L$ and the numerical resolution remains unchanged.  Nodes are placed at
$x_i=i\Delta x$, and strains and stresses are evaluated on bonds,
\begin{equation}
\begin{gathered}
    \varepsilon_{i+1/2}=\frac{u_{i+1}-u_i}{\Delta x},\\
    \sigma_{i+1/2}=Y\varepsilon_{i+1/2}+\beta_{\text{eff}}\varepsilon_{i+1/2}^3.
\end{gathered}
\end{equation}
Although this conservative discretization has a nearest-neighbor
algebraic structure, $\Delta x$ is a numerical resolution scale rather than a
microscopic lattice spacing. Rewriting the nodal kinetic and bond elastic
energies in the standard chain form gives the mesh-dependent coefficients
\begin{equation}
    m_h=\rho A\Delta x,\qquad
    K_h=\frac{AY}{\Delta x},\qquad
    \beta_h=\frac{A\beta_{\mathrm{eff}}}{\Delta x^3}.
    \label{eq:mesh-coefficients}
\end{equation}
Under mesh refinement these coefficients change while $\rho$, $Y$, and
$\beta_{\mathrm{eff}}$ remain fixed. The nearest-neighbor form therefore
expresses a flux-conservative numerical approximation of the continuum field,
not a reinterpretation of the mesh nodes as microscopic particles.
The acceleration is computed as a conservative stress divergence,
\begin{equation}
    a_i=\frac{\sigma_{i+1/2}-\sigma_{i-1/2}}{\rho\Delta x},\qquad 0<i<N-1,
\end{equation}
with free-end conditions $\sigma_{-1/2}=\sigma_{N-1/2}=0$, giving
\begin{equation}
\begin{gathered}
    a_0=\frac{\sigma_{1/2}}{\rho\Delta x},\\
    a_{N-1}=-\frac{\sigma_{N-3/2}}{\rho\Delta x}.
\end{gathered}
\end{equation}
The numerical scheme integrates the full nonlinear equations implied by the
constitutive stress, without a weak-anharmonic approximation.
The dynamics are advanced with a symmetric BAOAB splitting, combining
velocity-Verlet evolution of the conservative force with an exact
Ornstein--Uhlenbeck thermostat step \cite{leimkuhler2013}. The reservoirs act on $b=4$ nodes at
each end at the reference resolution. The corresponding values
for $\Delta x=1/64$ and $1/254$ are $b=2$ and $8$, respectively. With the
effective layer width defined as $b\Delta x$, these choices give
$0.03125$, $0.03150$, and $0.03150$ at the three resolutions, keeping the
thermostatted length approximately constant. Here $b$ counts nodes per
reservoir (Table~\ref{tab:numerical-parameters}); there are two reservoirs.
Keeping $b$ fixed would instead shrink the contact layer under refinement.
The choice thus preserves contact geometry rather than targeting a fitted
exponent. Matching $b\Delta x$ does not, by itself, establish equivalent
dynamical coupling to the reservoirs across resolutions.
For a thermostatted node coupled to reservoir $a\in\{h,c\}$,
\begin{equation}
    v_i\leftarrow e^{-\gamma\Delta t}v_i
    +\frac{\sqrt{T_a}}{\sqrt{\Delta x}}
    \sqrt{1-e^{-2\gamma\Delta t}}\,Z_{a,i},
\end{equation}
where the $Z_{a,i}$ are independent standard normal variables and
$\gamma=\gamma_0/(\rho\Delta x)$. At the reference resolution,
the simulations use
$\gamma\Delta t=0.005$, and the bath parameters are held fixed as $L$ is
varied.  The initial velocities are sampled from a linear temperature profile
between the two reservoirs.  After equilibration, observables are averaged in
the nonequilibrium steady state over $M=128$ independent realizations and
over every production step. The reference runs use the size-dependent
thermalization and production intervals listed in Table~\ref{tab:numerical-parameters}.
The approach to a stationary state is monitored through the nearly
position-independent time-averaged bulk current and the stationary temperature
profiles reported below.
To control drift of the translational zero mode, the mean
displacement and center-of-mass velocity are subtracted every 1000 steps.
Subtracting the displacement leaves all strains unchanged. The velocity
projection $v_i'=v_i-V_{\mathrm{cm}}$ also preserves the instantaneous
relative velocities, $v_{i+1}'-v_i'=v_{i+1}-v_i$. It sets the total momentum
to zero and removes only the kinetic energy of the collective translation,
\begin{equation}
    E_{\mathrm{proj}}=\frac{Nm}{2}V_{\mathrm{cm}}^2,\qquad
    V_{\mathrm{cm}}=\frac{1}{N}\sum_{i=0}^{N-1}v_i,
    \label{eq:projection-energy}
\end{equation}
where $m=\rho A\Delta x$ is the nodal mass and $V_{\mathrm{cm}}$ is evaluated
immediately before each projection. This preservation of the instantaneous
internal motion does not establish invariance of the subsequent thermostatted
dynamics, because the Langevin forces act on the nodal velocities. The effect
of the projection on transport has not been isolated quantitatively.

The continuous energy current associated with Eq.~\eqref{eq:bulk} is
\begin{equation}
    j(x,t)=-v(x,t)\sigma(\partial_x u).
\end{equation}
On the grid we measure the bond current
\begin{equation}
    j_{i+1/2}=-\frac{1}{2}\left(v_i+v_{i+1}\right)\sigma_{i+1/2}.
\end{equation}
For the conservative semidiscrete dynamics, define the nodal energy
\begin{equation}
    e_i=\frac{m v_i^2}{2}+\frac{A\Delta x}{2}
    \left[W(\varepsilon_{i-1/2})+W(\varepsilon_{i+1/2})\right].
\end{equation}
The stress divergence and the bond current above then obey the exact
semidiscrete identity
\begin{equation}
    \frac{d e_i}{dt}+A\left(j_{i+1/2}-j_{i-1/2}\right)=0,
    \qquad 0<i<N-1.
    \label{eq:discrete-energy-balance}
\end{equation}
Indeed, using
$\dot W(\varepsilon_{i+1/2})=\sigma_{i+1/2}
\dot\varepsilon_{i+1/2}$ and
$\dot\varepsilon_{i+1/2}=(v_{i+1}-v_i)/\Delta x$, the stress-work terms in
$d e_i/dt$ combine exactly into the difference of the two adjacent bond
currents. Summation over any contiguous set of interior nodes therefore
telescopes to its two boundary currents. Thus, the spatial discretization does
not introduce a residual bulk source or sink of energy.
This identity concerns the conservative semidiscrete equations,
not the thermostat updates or velocity projections. The full numerical energy
budget also includes Langevin exchange, the energy removed by
Eq.~\eqref{eq:projection-energy}, and finite-time-step integration errors.
The current is averaged in time and over the interior
of the chain. More
specifically, with
$i_{\mathrm{cut}}=\max(b+2,\lfloor N/32\rfloor)$, we define
\begin{equation}
    J_{\mathrm{bulk}}=
    \frac{1}{N-2i_{\mathrm{cut}}}
    \sum_{i=i_{\mathrm{cut}}}^{N-1-i_{\mathrm{cut}}}J_i,
    \qquad J_i=\frac{j_{i-1/2}+j_{i+1/2}}{2},
    \label{eq:bulk-current}
\end{equation}
and use the corresponding interior temperature drop to obtain
\begin{equation}
    \kappa=\frac{J_{\mathrm{bulk}}L_{\mathrm{bulk}}}
    {\overline T_{i_{\mathrm{cut}}}
    -\overline T_{N-1-i_{\mathrm{cut}}}},
    \qquad L_{\mathrm{bulk}}=(N-1-2i_{\mathrm{cut}})\Delta x,
    \label{eq:kappa-estimator}
\end{equation}
where $\overline T_i$ denotes the time- and ensemble-averaged kinetic
temperature. This definition excludes the thermostatted layers and their
immediate contact region while retaining an extensive bulk interval.
The contact exclusion is specified before the conductivity fit and
scales as a fixed fraction of the chain once $N/32>b+2$; it is therefore not
adjusted independently for each system size to optimize the exponent.

\section{Results and discussion}

\subsection{Temperature profile and contact effects}

The conservative discretization makes the measured current consistent with
the stress divergence used in the equations of motion.  In a nonequilibrium
steady state, the time-averaged bond current should become approximately
independent of position away from the thermostatted boundaries.  With the
lumped nodal mass $m=\rho A\Delta x$, equipartition defines the local kinetic
temperature as
\begin{equation}
    k_{\mathrm B}\overline T_i=m\langle v_i^2\rangle.
\end{equation}
Thus, in the reduced units used in the simulations,
$\overline T_i=\Delta x\langle v_i^2\rangle$.  This normalization is consistent with
the factor $1/\sqrt{\Delta x}$ in the thermostat noise amplitude and permits a
direct comparison with the imposed reservoir temperatures.  The resulting
stationary temperature profiles are shown in
Fig.~\ref{fig:temperature-profile}.

The near-collapse of the profiles in Fig.~\ref{fig:temperature-profile}
indicates comparable nonequilibrium steady states for the larger systems. The
steeper variation near the ends reflects
reservoir contact effects and should not be interpreted as a bulk thermal
gradient; conductivity estimates are consequently based on the interior of
the chain. To quantify the profile collapse, let
$\xi=x/L$ and take $\overline{T}_{\mathrm{ref}}(\xi)$ to be the stationary
profile for $L_{\max}=64.50$. For each length $L$,
$\overline{T}_L(\xi)$ is linearly interpolated onto a common uniform grid of
$P$ points over the bulk interval $0.1\leq\xi\leq0.9$. We then define the
root-mean-square collapse distance
\begin{equation*}
    \Delta_T(L)=
    \left[\frac{1}{0.8}
    \int_{0.1}^{0.9}
    \left(\overline{T}_L(\xi)-
    \overline{T}_{\mathrm{ref}}(\xi)\right)^2d\xi\right]^{1/2}.
\end{equation*}
In practice, the integral is evaluated by the corresponding
average over the $P$ interpolated points. The inset of
Fig.~\ref{fig:temperature-profile} shows a rapid decrease of $\Delta_T(L)$
through the short-length range and a residual level
$\Delta_T\simeq0.32$ for $L>8$. Relative to the imposed temperature
difference $T_h-T_c=85$, this corresponds to less than $0.4\%$ and supports
the use of this interval in the subsequent flux and conductivity analyses.

\subsection{Flux collapse and conductivity scaling}

\begin{figure}[!htbp]
    \centering
    \includegraphics[width=\columnwidth]{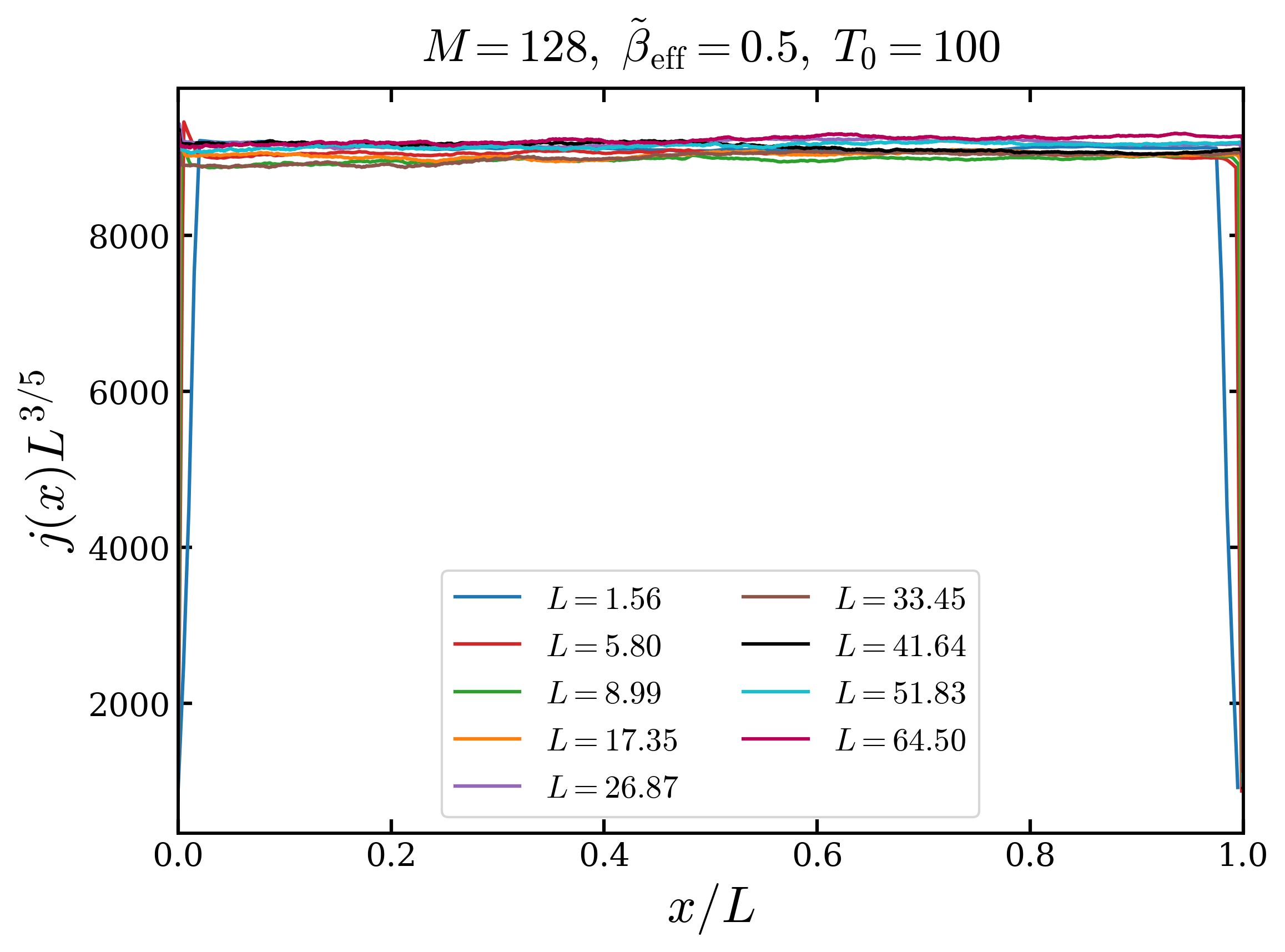}
    \caption{Rescaled stationary flux density $j(x)L^{3/5}$ versus $x/L$.
    The broad, nearly horizontal plateau is consistent with a stationary bulk
    energy balance: the mean current is approximately independent of bond
    index.
    The collapse for different $L$ supports
    $j\sim L^{-3/5}$, whereas the sharp endpoint deviations are produced by
    the Langevin contacts and the definition of the boundary bond current.}
    \label{fig:rescaled-flux}
\end{figure}

Figure~\ref{fig:rescaled-flux} tests the predicted finite-size scaling of the
stationary current by plotting the rescaled flux across the chain.

\begin{figure}[!htbp]
    \centering
    \includegraphics[width=\columnwidth]{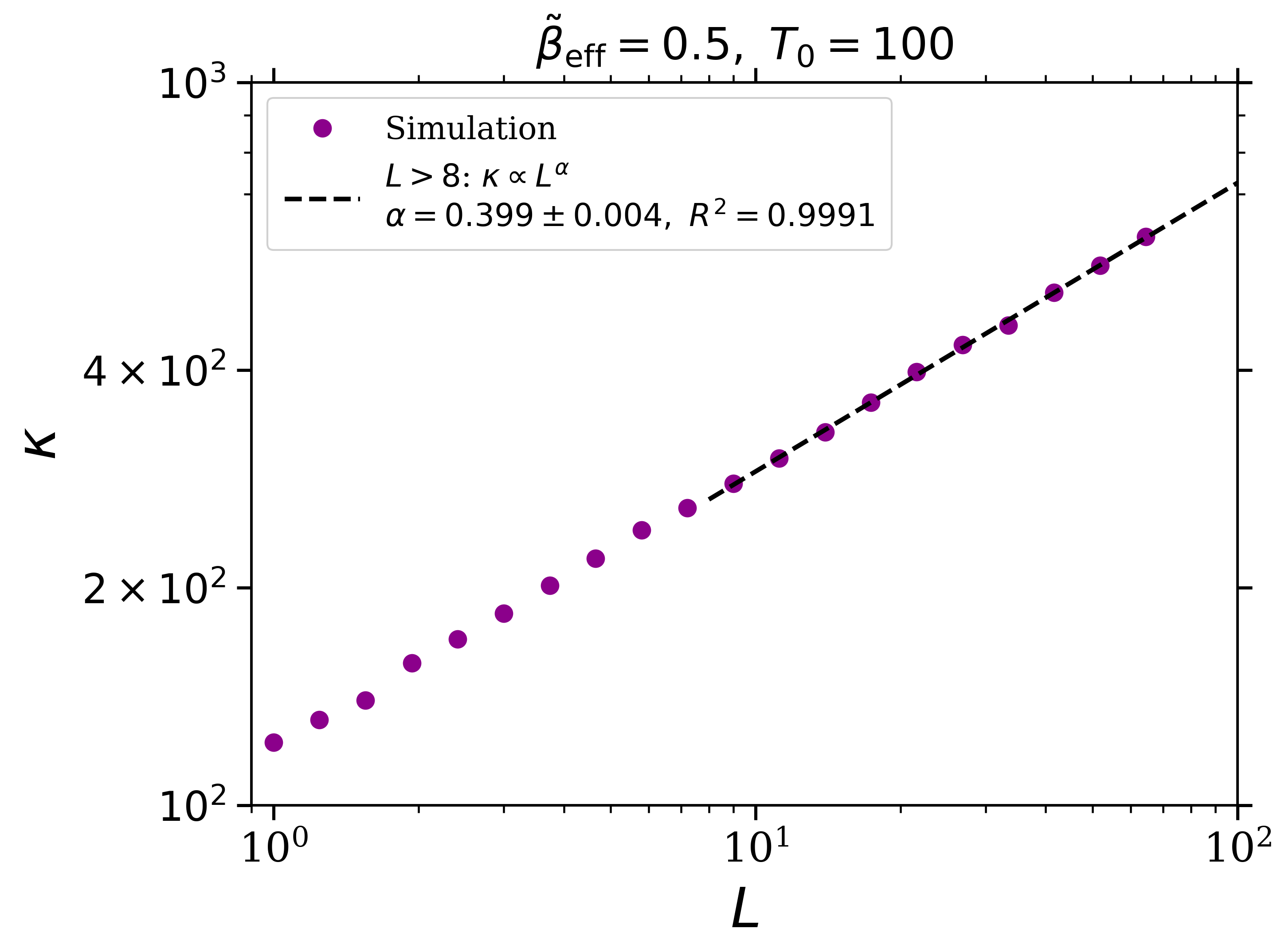}
    \caption{Effective thermal conductivity versus system length on logarithmic
    axes.  For $L>8$, the fit gives
    $\kappa\propto L^{\alpha}$ with $\alpha=0.399\pm0.004$, consistent with
    the $2/5$ kinetic-theory benchmark. The growth of
    $\kappa$ with $L$, rather than saturation to a constant, is the defining
    signature of anomalous conduction and is compatible with the long-lived
    collective modes of the momentum-conserving $\beta$-FPUT chain.}
    \label{fig:kappa}
\end{figure}

As shown in Fig.~\ref{fig:rescaled-flux}, the curves collapse onto a broad
bulk plateau. In the reduced units used here and at fixed reservoir temperature
difference, Fourier's law
would instead give $j\propto L^{-1}$ with a size-independent conductivity.  The
slower decay $j\propto L^{-3/5}$ is therefore compatible with anomalous heat
conduction in a momentum-conserving $\beta$-FPUT chain. Since the conductivity
estimator in Eq.~\eqref{eq:kappa-estimator} uses the interior temperature drop,
its size dependence is assessed directly in Fig.~\ref{fig:kappa}.

\begin{figure}[!htbp]
    \centering
    \includegraphics[width=\columnwidth]{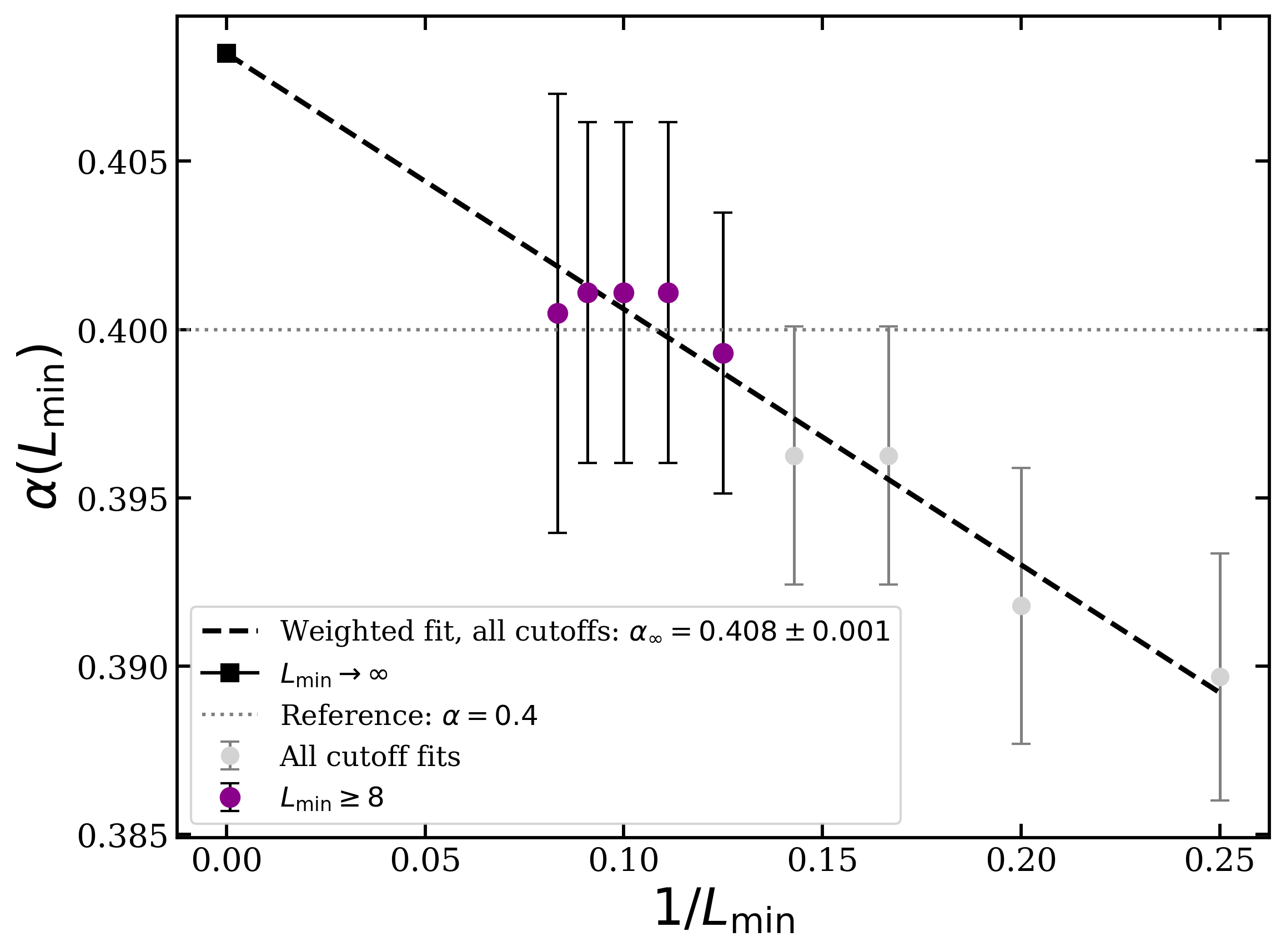}
    \caption{Systematic cutoff analysis of the conductivity
    exponent. Each point is obtained from a log--log fit of $\kappa$ versus $L$ retaining data with $L\ge L_{\min}$, for $L_{\min}=4,\ldots,12$; error bars are the standard errors of the individual fitted slopes.
    Purple symbols highlight the fits with $L_{\min}\ge8$, while gray symbols denote the full set of cutoff fits. The horizontal dotted line is the reference value $\alpha=0.399$ from the $L>8$ fit in Fig.~\ref{fig:kappa}. The dashed line is a weighted linear regression of all cutoff estimates versus $1/L_{\min}$, with empirical-fit intercept $\alpha_\infty=0.408\pm0.001$.}
    \label{fig:alpha-cutoff}
\end{figure}

The agreement between the flux collapse in Fig.~\ref{fig:rescaled-flux} and
the conductivity fit in Fig.~\ref{fig:kappa} supports the internal
consistency of the simulation. The shortest chains are excluded from the
power-law fit because they show stronger finite-size and contact effects.
The cutoff $L>8$ is selected
because the profile metric in Fig.~\ref{fig:temperature-profile} and the
rescaled current in Fig.~\ref{fig:rescaled-flux} identify the same range.
An ordinary least-squares fit of
$\log\kappa$ against $\log L$ for $L>8$ therefore tests the conductivity
growth in the range where the current-collapse criterion is satisfied. The fit
gives the slope quoted in Fig.~\ref{fig:kappa}; the reported uncertainty is
its one-standard-error regression estimate.
This error does not explicitly propagate the sampling uncertainty
of each conductivity estimate or account for systematic effects from the
fitting range and numerical protocol.
The fitted value is consistent
with $L^{2/5}$ behavior over the selected size range. For comparison, fitting all
available lengths gives $\alpha=0.387\pm0.002$. This value remains anomalous,
but it mixes the short-length crossover and contact-dominated points with the
range selected by the current-collapse criterion. This comparison is controlled because
$\Delta x$, $\Delta t$, the reservoir temperatures, and the thermostat
coupling $\gamma_0$ are held fixed while only the number of bonds is varied.
At $\Delta x=1/127$, the threshold $L>8$ corresponds to
$N>1017$, or approximately $10^3$ mesh nodes. The effective exponent therefore
stabilizes near $2/5$ with a moderate number of continuum degrees of freedom,
whereas direct simulations of the microscopic chain have been extended far
beyond $N=10^5$ to investigate their long finite-size crossover
\cite{takatsu2024,lepri2026}.
This comparison motivates the practical use of the field
formulation, but it is not a matched-parameter benchmark: a difference in
node counts alone does not establish a computational speedup or identify
the cause of the shorter observed crossover.
Accordingly, the evidence for the selected range is joint rather
than based on the conductivity fit alone: the same cutoff is identified from
the spatial structure of the rescaled current, and the all-length
fit demonstrates that including the noncollapsed short chains shifts the
effective exponent downward. These diagnostics are complementary,
not statistically independent, since the conductivity estimator uses the
same current data. Figure~\ref{fig:alpha-cutoff} makes this fitting
range dependence explicit. The individual estimates for $L_{\min}\ge8$ remain
close to the reference value, whereas the empirical-fit intercept
$\alpha_\infty=0.408\pm0.001$ is modestly higher. The cutoff analysis
therefore provides an additional view of the finite-size trend.
The cutoff fits share most of their input data, so their slope
estimates are correlated. The uncertainty of the weighted linear-fit
intercept should therefore be read as a formal regression error, not as a
complete uncertainty on a thermodynamic-limit exponent. The linear dependence
on $1/L_{\min}$ is empirical and does not establish the asymptotic correction
law.

\subsection{{Sensitivity to spatial resolution}}
\label{sec:mesh-refinement}

\begin{figure}[!t]
    \centering
    \includegraphics[width=0.95\textwidth]{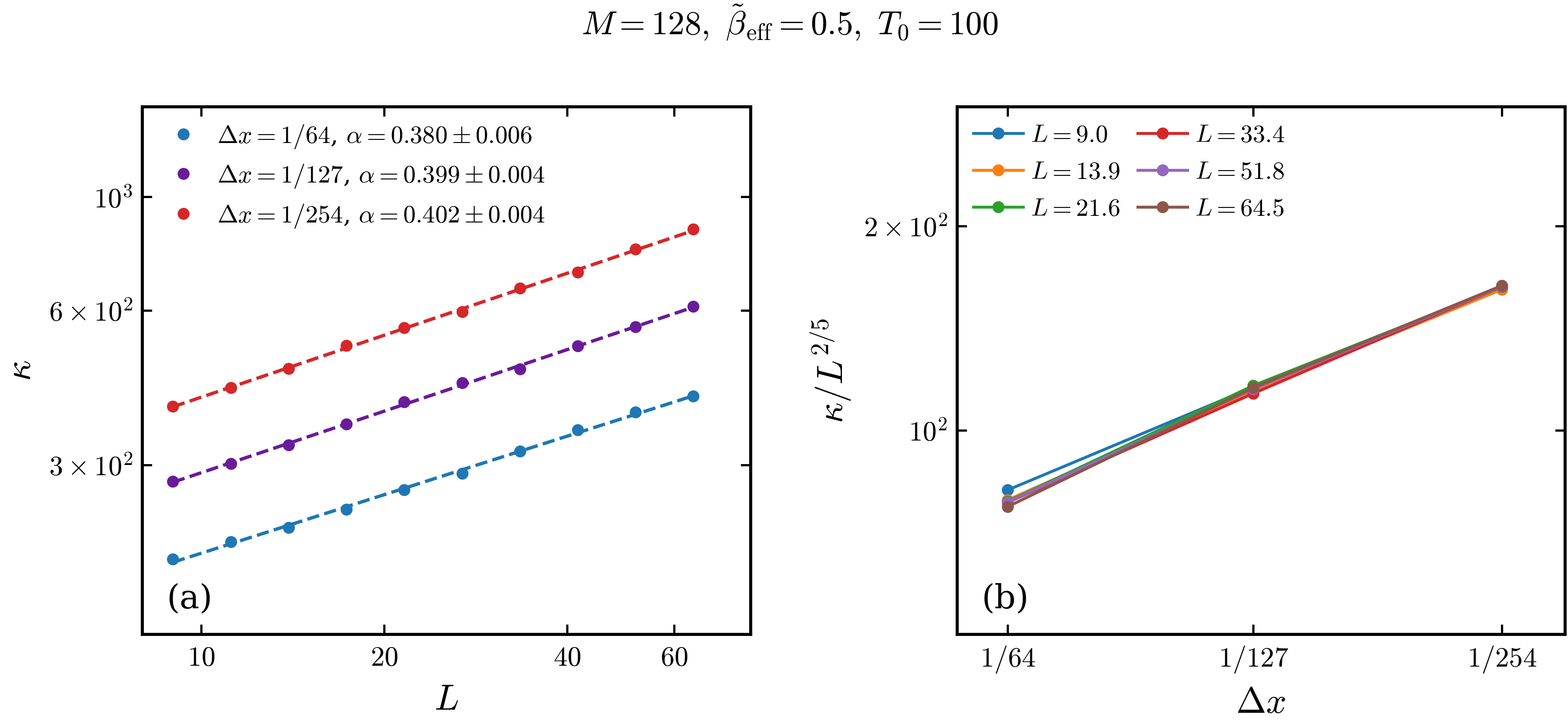}
    \caption{{Mesh-refinement test of anomalous thermal transport
    for $M=128$, $\widetilde{\beta}_{\mathrm{eff}}=0.5$, and $T_0=100$.
    The physical width of the thermal reservoirs is kept approximately
    constant across the three resolutions. (a) Bulk thermal conductivity as
    a function of physical chain length for $\Delta x=1/64$, $1/127$, and
    $1/254$. Symbols denote simulation data and dashed lines are power-law
    fits, $\kappa=C(\Delta x)L^\alpha$, over $L\geq8.9$. The two finer
    resolutions give exponents consistent with $2/5$; the coarsest mesh gives
    a lower effective exponent. (b) Rescaled conductivity $\kappa/L^{2/5}$
    at selected matched physical lengths. The prefactor remains dependent
    on spatial resolution over the tested range. Lines connect points to
    guide the eye.}}
    \label{fig:mesh-refinement}
\end{figure}

To assess the sensitivity of the anomalous conductivity scaling to
the spatial discretization, we repeated the simulations at
$\Delta x=1/64$, $1/127$, and $1/254$, keeping the physical width of the
thermal reservoirs approximately fixed. The comparison uses $M=128$,
$\widetilde{\beta}_{\mathrm{eff}}=0.5$, and mean reservoir temperature
$T_0=(T_h+T_c)/2=100$. As shown in Fig.~\ref{fig:mesh-refinement}(a), all
three resolutions exhibit a power-law increase of the bulk conductivity with
physical length. Fits of $\kappa=C(\Delta x)L^\alpha$ use the physical
length $L=(N-1)\Delta x$ and ten nearly matched lengths over $L\geq8.9$, giving
$\alpha=0.380\pm0.006$, $0.399\pm0.004$, and $0.402\pm0.004$,
respectively. The two finer meshes agree within the quoted uncertainties and
are consistent with $\alpha=2/5$, whereas the coarsest mesh yields a lower
effective exponent. Thus, the comparison supports the stability of the
exponent near $2/5$ between the two finer resolutions, while revealing
residual spatial-resolution effects on the coarsest mesh.

Reanalysis of the existing profiles provides additional tests
of this scaling. Restricting all three data sets to $L\geq13.9$ retains
eight lengths per resolution and gives $\alpha=0.3926\pm0.0057$,
$0.4005\pm0.0065$, and $0.4023\pm0.0065$, respectively. The reduced
discrepancy of the coarsest mesh is consistent with a stronger finite-size
crossover at that resolution. Separately, omitting one length at a time from
the ten-point fits gives exponent ranges of $0.3979$--$0.4011$ for
$\Delta x=1/127$ and $0.3998$--$0.4040$ for $1/254$, indicating that no
single length determines the fine-mesh results. These are sensitivity tests;
the primary fitting interval remains $L\geq8.9$.

The interior estimator was also tested with
$i_{\mathrm{cut}}=\max(b+2,\lfloor fN\rfloor)$ for
$f=1/32$, $0.05$, $0.10$, $0.15$, and $0.20$, retaining the ten lengths
at each resolution. The resulting exponent ranges are
$0.3752$--$0.3847$, $0.3993$--$0.4097$, and $0.4020$--$0.4129$
for $\Delta x=1/64$, $1/127$, and $1/254$, respectively. The anomalous
growth persists, but the fitted exponent depends moderately on the portion
of the temperature profile used to define the interior drop. These ranges
describe estimator sensitivity, not statistical confidence intervals.

The mesh spacing nevertheless affects the amplitude of the
conductivity. Figure~\ref{fig:mesh-refinement}(b) isolates this effect by
plotting $\kappa/L^{2/5}$ at matched physical lengths: the rescaled values
increase systematically as the mesh is refined, without a visible prefactor
plateau in the present range. This is consistent with a residual dependence
of the conductivity amplitude on the ultraviolet cutoff introduced by the
spatial mesh. Attributing the entire variation to this cutoff would also
require controlling the reservoir coupling and the other resolution-dependent
numerical operations. The distinction is important: stability of the size-scaling
exponent does not imply convergence of the conductivity itself as
$\Delta x\to0$. The present tests support a robust exponent near $2/5$ at
the finer resolutions, but do not establish a cutoff-independent conductivity
amplitude.

\section{Final remarks}

The central result of this work is that a long-wave continuum
formulation of the symmetric $\beta$-FPUT chain exhibits a conductivity
exponent close to $2/5$ at accessible numerical sizes. For $L>8$, or
approximately $10^3$ mesh nodes and above at the reference resolution, we
obtain $\alpha=0.399\pm0.004$ (regression standard error). The temperature
profiles and rescaled-current collapse provide complementary evidence for
the same finite-size regime. This addresses the practical motivation for
the continuum approach: accessing anomalous scaling without resolving the
very large microscopic chains used to investigate long crossovers in
molecular dynamics. Such studies have explored particle numbers well beyond
$10^5$ \cite{takatsu2024,lepri2026}. The present result demonstrates access
to the scaling regime with a moderate number of field variables; a measured
computational speedup over molecular dynamics would require a benchmark with
matched physical parameters, reservoir protocols, and statistical accuracy.

The distinction between the continuum model and its numerical
mesh is essential to this interpretation. The nodes represent the
displacement field, rather than individual microscopic particles. Both
forces and energy currents follow from the same nonlinear stress, and the
exact conservative semidiscrete continuity identity ensures a consistent
local energy balance. This makes the formulation useful for investigating
how anharmonic elasticity and thermal contacts shape energy transport,
with an interior estimator that excludes the directly thermostatted regions.

The mesh-refinement results distinguish two aspects of transport:
the size-scaling exponent is stable between $\Delta x=1/127$ and $1/254$,
whereas the conductivity amplitude retains a measurable resolution
dependence. The coarser-mesh exponent approaches $2/5$ when the two shortest
lengths are excluded, and the fine-mesh estimates are insensitive to omission
of any single length. These observations support the use of the formulation
to study anomalous size scaling, without establishing a cutoff-independent
absolute conductivity. The interior-window tests also reveal estimator
sensitivity. Full sampling uncertainties and the influence of the velocity
projection remain to be quantified; the quoted regression errors do not
include these contributions.

Beyond the statistical-mechanical setting, the continuum
description provides a connection to engineering models of nonlinear
elasticity: its constitutive stress derives from a strain-energy density,
and its numerical energy flux is consistent with that constitutive law.
This structure offers a starting point for extensions to nonlinear elastic
waveguides and studies of how strain-dependent stiffness affects wave
propagation and energy transfer \cite{barth2024,zhang2024}. Such extensions
would require the appropriate geometry and material calibration, and
viscoelastic or other dissipative variables when damping is important;
quantitative predictions for engineered materials are not established by the
present transport calculations. Adjacent engineering examples
include pre-stress-programmable mechanical metamaterials designed to tune
energy dissipation and high-damping rubber composites
\cite{wang2025,zhao2015}. The contribution is therefore both a
numerically accessible testbed for anomalous transport in anharmonic systems
and a conservative basis for developing models of energy flow in nonlinear
elastic media.

\section*{Acknowledgments}
We acknowledge fruitful discussions and support from Rodrigo C. V. Coelho. We also acknowledge financial support from CNPq and CAPES (Brazilian agencies). We further acknowledge CBPF for computational resources provided via the HPC Sci.Mind infrastructure of LAB-IA (Laboratory of Artificial Intelligence), affiliated with COTEC, where the simulations reported in this work were performed.

\end{document}